\documentstyle[sprocl,amssymb,amsmath,epsfig,calc]{article}

\def\Journal#1#2#3#4{{#1} {\bf #2}, #3 (#4)}

\def\aa{\em A.\& A.}
\def\apj{\em Ap.J.}

\def\mnras{\em M.N.R.A.S.}

\def\be{\begin{equation}}
\def\ee{\end{equation}}
\def\bea{\begin{eqnarray}}
\def\eea{\end{eqnarray}}

\begin{document}
\title{CHARACTERIZING CLUSTER MORPHOLOGY  USING VECTOR-VALUED MINKOWSKI FUNCTIONALS}

\author{ C. BEISBART and T. BUCHERT }

\address{Theoretische Physik, Ludwig-Maximilians-Universit\"at,\\ Theresienstr. 37, D-80333 M\"unchen, Germany\\emails:\;\;beisbart,buchert@stat.physik.uni-muenchen.de}


\maketitle\abstracts{
The morphology of galaxy clusters is quantified using Minkowski functionals, especially the vector--valued ones, which contain directional information and are related to curvature centroids. The asymmetry of clusters and the amount of their substructure can be characterized in a unique way using these measures. -- We briefly introduce vector--valued Minkowski functionals (also known as Querma\ss\;vectors) and suggest their application to cluster data in terms of a morphological characterization of excursion sets. Furthermore, we develop robust structure functions which describe the dynamical state of a cluster and study the evolution of clusters using numerical simulations.
}
\section{Introduction}
The substructure of galaxy clusters is important not only for its own sake. For the usual mass estimates of clusters depend on how the mass distribution is modelled, especially whether substucture is ignored or not. Furthermore, the amount of substructure observed in clusters can serve as a sensitive probe of the cosmological parameters, especially the overall matter density $\Omega_0$~\cite{evr}.\\
In this context it is necessary to measure cluster substructure in a quantitative way. So far, only very crude substructure measures have been developed (for an overview see Crone et al.~\cite{crone}, for a different approach Grebenev et al.~\cite{grebenev}). In our contribution, we use Minkowski functionals to characterize cluster morphology in the systematic framework of integral geometry. The Minkowski functionals can be introduced in an axiomatic way (Section \ref{sec:2}) and have been sucessfully applied in discriminating large scale structure~\cite{aco,fluctuations}. We focus on vector--valued Minkowski functionals which also incorporate directional information.
\section{The Minkowski functionals}\label{sec:2}
\begin{figure}
\begin{center}
\epsfig{figure=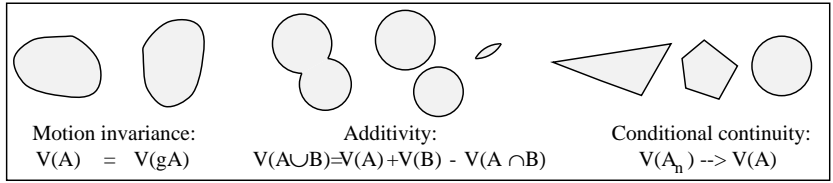,height=2.4cm}
\caption{The requirements defining the Minkowski functionals}
\label{fig:minkf}
\end{center}
\end{figure}

\begin{figure}[b]
\centering
\epsfig{figure=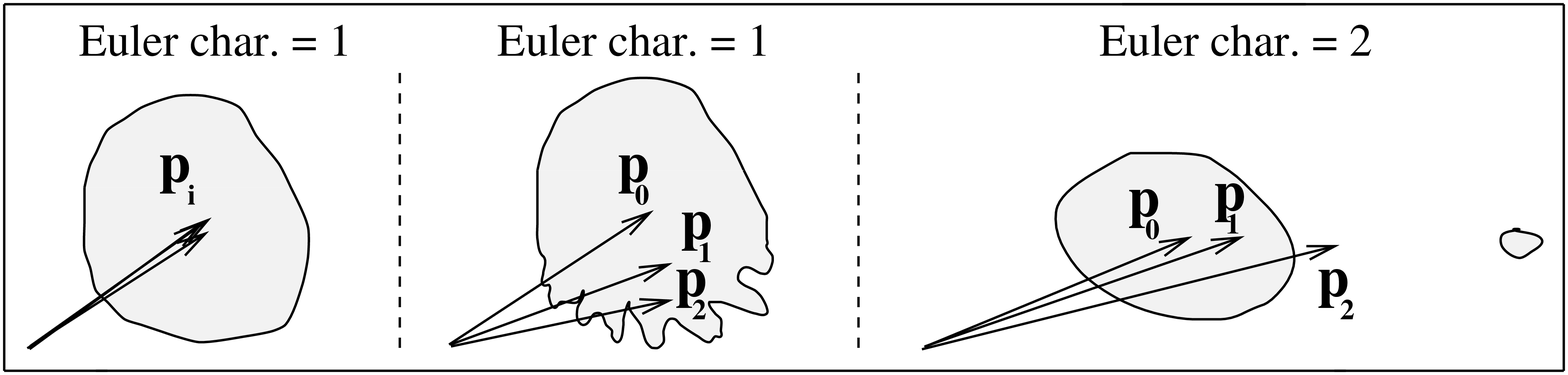,height=2.4cm}
\caption{Examples}
\label{fig:examples}
\end{figure}

The (scalar) Minkowski functionals are defined for closed bodies and can be characterized by simple requirements such as motion invariance, addivity and conditional continuity (see Figure \ref{fig:minkf}). Although these requirements are fairly general, Hadwiger's theorem~\cite{mecke} states that in $d$ dimensions there are only $d+1$ linear independent Minkowski functionals $V_i$. This means that a characterization of bodies is possible in a unique way in terms of $d+1$ measures which carry intuitive meanings. For example, in two dimensions, the first Minkowski functional, called $V_0$, is simply the surface content. The other two Minkowski functionals are the length of the circumference and the Euler characteristic; they can be obtained by integrating over the Lebesgues surface of the body and weighting with curvature measures. The Euler characteristic $V_2$ is related to the genus, for further details see Mecke et al.~\cite{mecke}.\\
An interesting extension can be made by going to vector--valued Minkowski functionals. These transform like vectors (motion {\em equi}variance), and it turns out that to each scalar Minkowski funtional $V_i$ there corresponds a vector--valued brother $\mathbf{V}_i$ or equivalently a  normalized (curvature) centroid $\begin{mathbf}p\end{mathbf}_i \equiv \frac{\begin{mathbf}V\end{mathbf}_i}{V_i}$, which is simply a point within the convex covering of the body, for further details compare Beisbart et al.~\cite{vec}. In Figure \ref{fig:examples} some examples are given how Querma\ss\;vectors work. Most notably, the (curvature) centroids are sensitive to symmetry. For a roughly symmetric body the centroids coincide, while they split up for an asymmetric configuration.
\section{The method: Minkowski functionals of excursions sets}\label{sec:3}
In order to employ the Minkowski functionals for the quantification of cluster morphology we have to associate bodies with the usual data sets (2d galaxy/X-ray-photon positions). For this purpose we smooth the galaxy positions with a Gaussian kernel to obtain a density field $u(\mathbf{x})$. The smoothing length serves as a dia\-gnostic parameter determining at which scale substructure is resolved. The morpho\-logy of the excursion sets $M_\nu = \left\{ \begin{mathbf}x\end{mathbf} \in  \begin{mathbb}R\end{mathbb}^{d}\mid\; u (\begin{mathbf}x\end{mathbf}) \geq \nu \right\}$ can now be described in terms of the Minkowski functionals~\cite{beyond}. For spherically symmetric reference models the excursions sets become simply circles, and the curvature centroids ${\mathbf p}_i$ coincide.\\
\begin{figure}
\begin{minipage}[t]{.24\linewidth}
\centering
\epsfig{figure=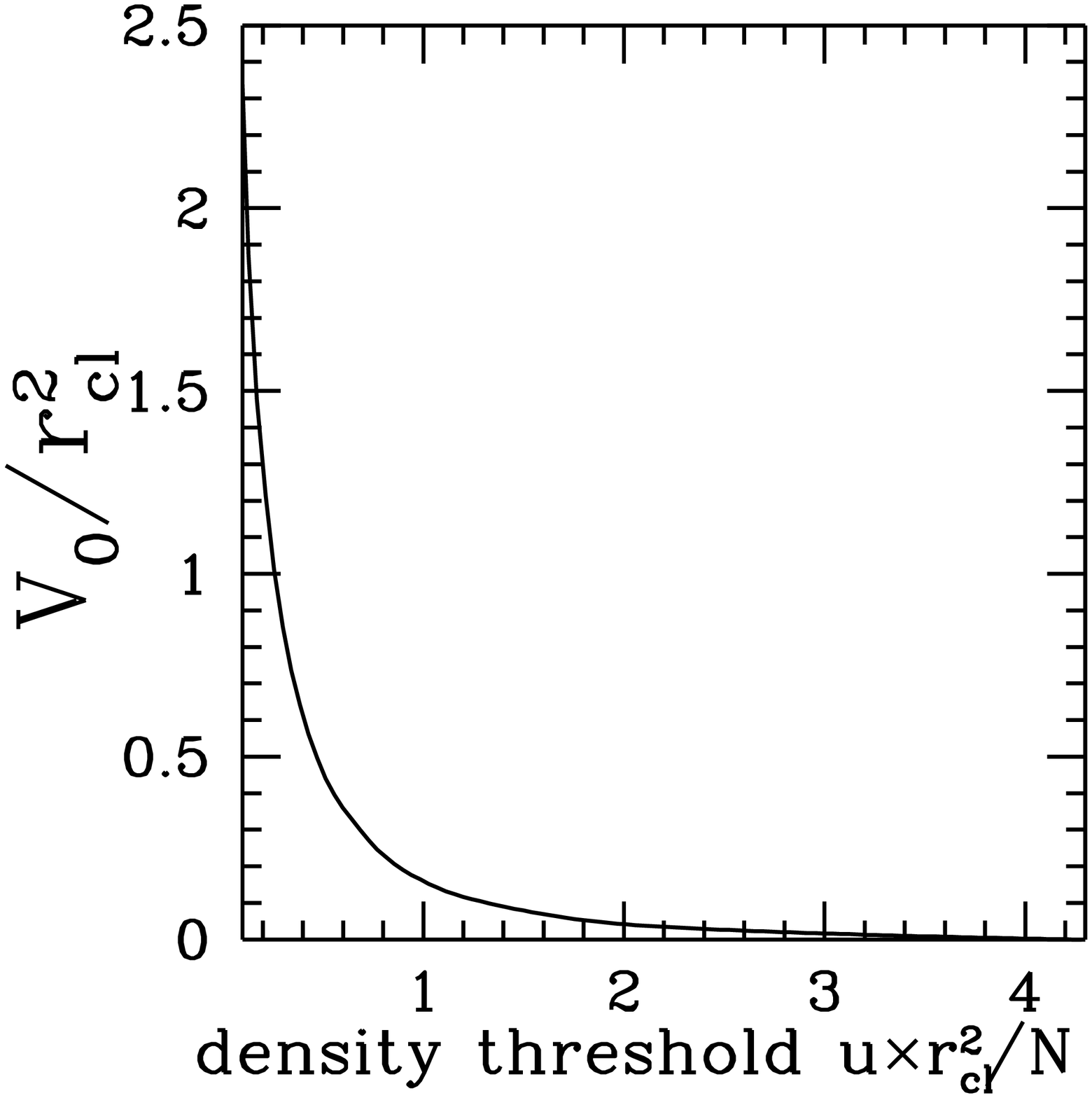,height=2.6cm,width=2.8cm}
\end{minipage}
\begin{minipage}[t]{.24\linewidth}
\centering
\epsfig{figure=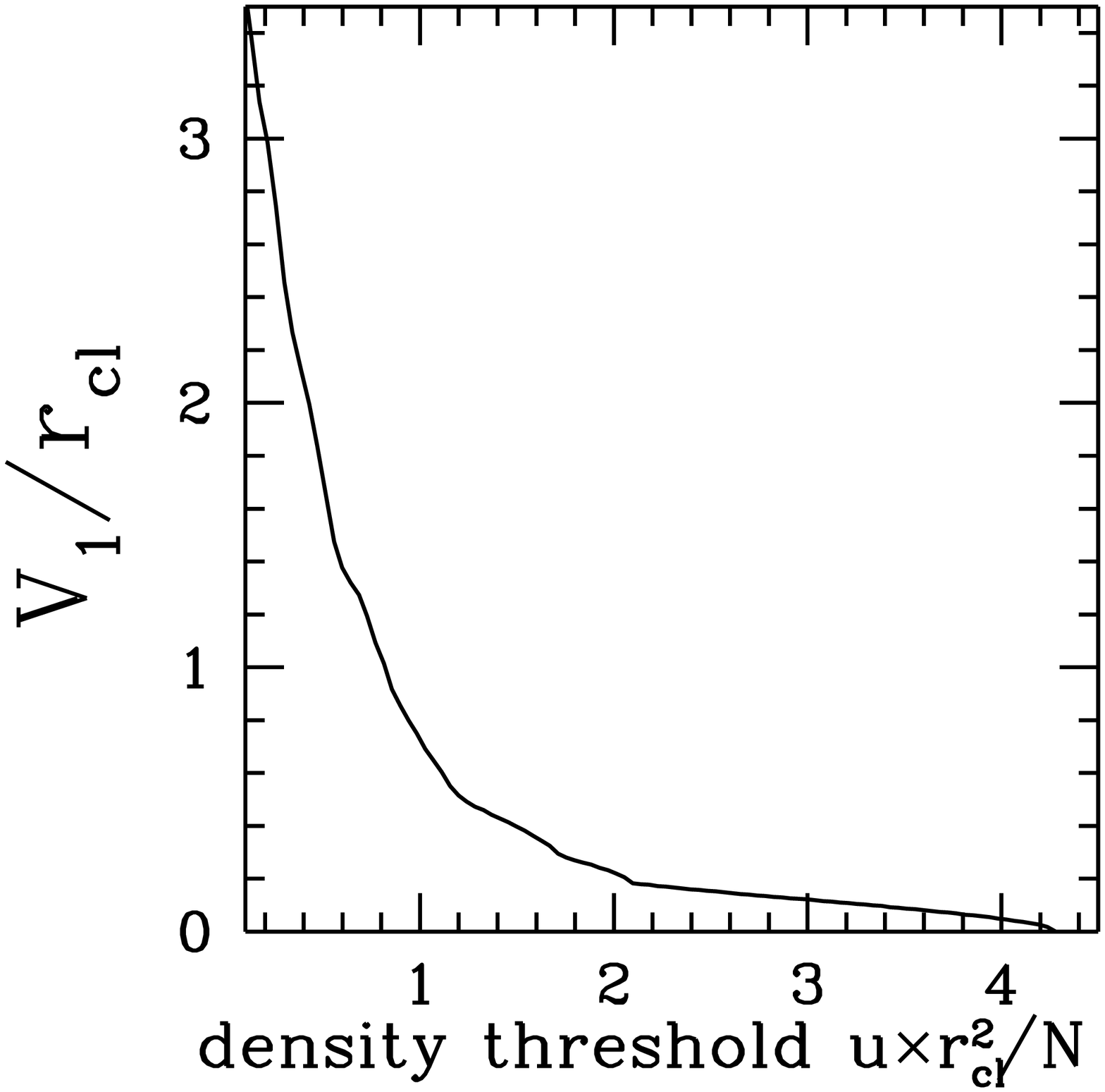,height=2.6cm,width=2.8cm}
\end{minipage}
\begin{minipage}[t]{.24\linewidth}
\centering
\epsfig{figure=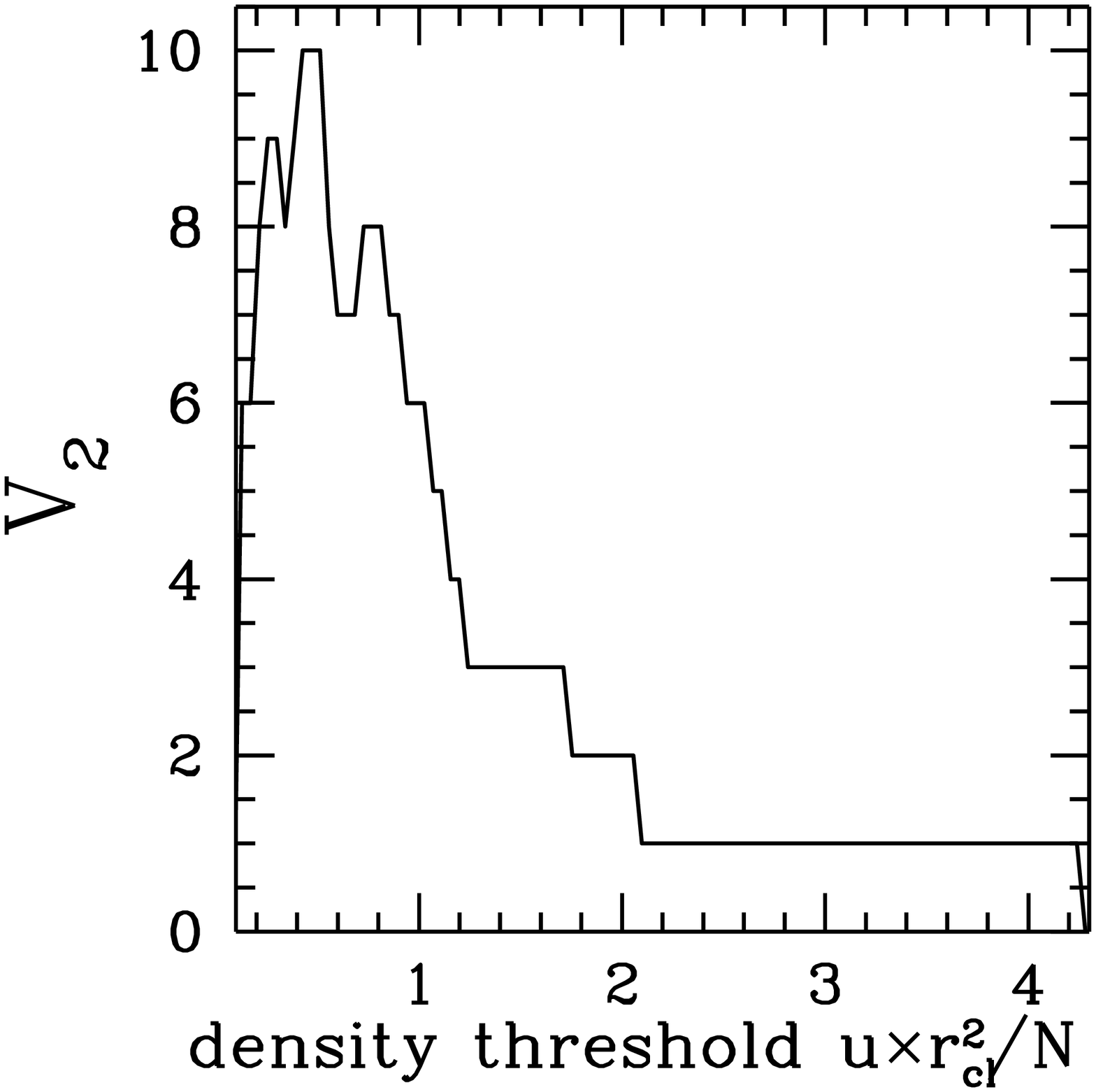,height=2.6cm,width=2.8cm}
\end{minipage}
\begin{minipage}[t]{.24\linewidth}
\centering
\epsfig{figure=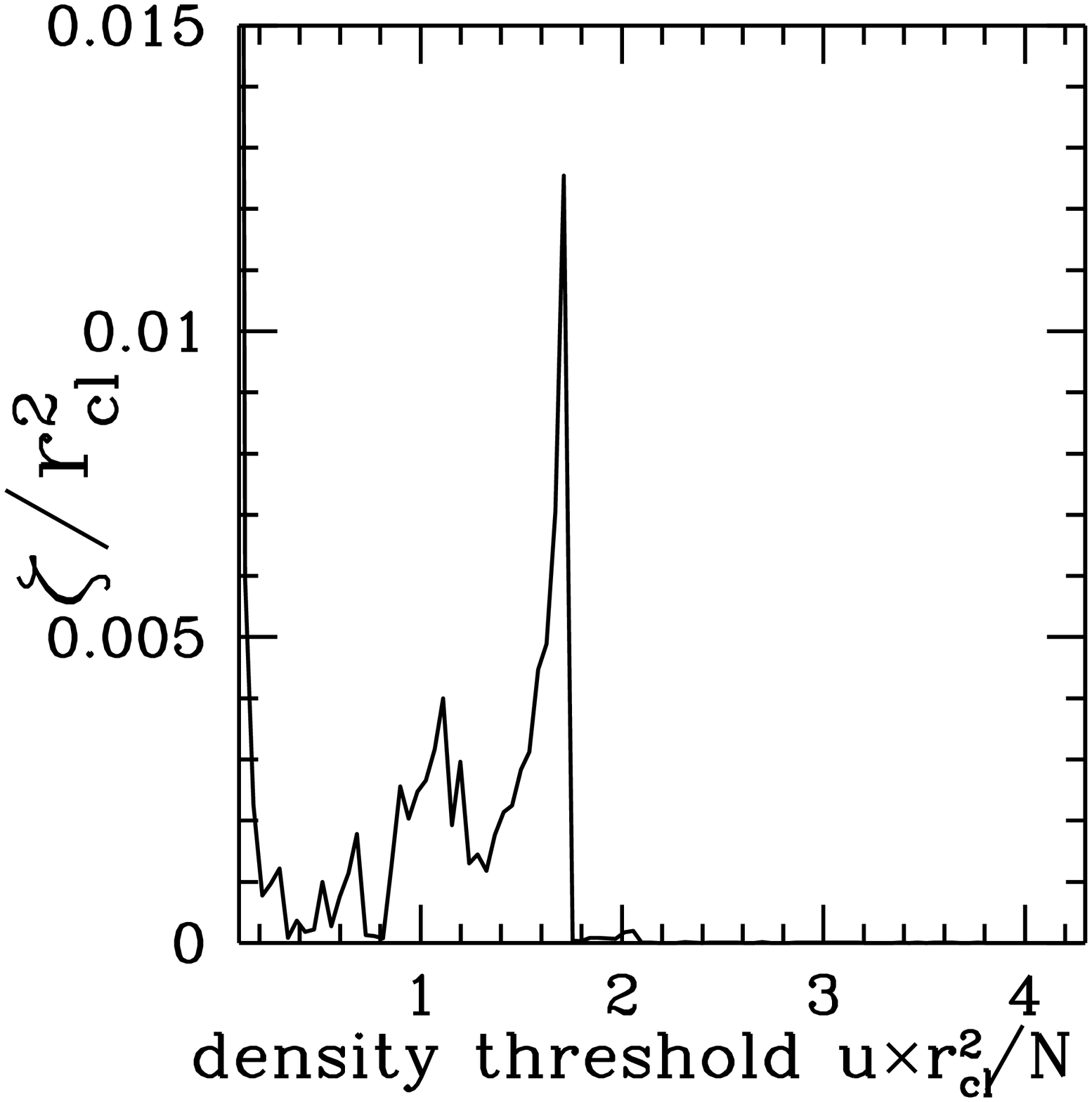,height=2.6cm,width=2.8cm}
\end{minipage}
\caption{The Minkowski functionals $V_0$, $V_1$, $V_2$ and the morphological parameter $\zeta$ are shown as functions of the density threshold $u$.}
\label{fig:minkclus}
\end{figure}
This changes if we proceed to analyzing more realistic data. For example, we investigate clusters simulated by Matthias Steinmetz for an $\Omega_0 = 1$--universe~\cite{stein}. In Figure \ref{fig:minkclus} the behaviour of the scalar Minkowski functionals $V_i$ is shown in dependence of the density threshold defining the excursion set; so we get individual signatures of the cluster: the Euler characteristic $V_2$ simply counts the components of the cluster; for high density thresholds only the core of the cluster remains (resulting in $V_2=1$), whereas for lower levels up to 10 components are counted; the se\-cond Minkowski functional $V_1$ (second panel), the length of the circumference, gives an averaged, smoothed and inversed density profile of the cluster. The ratio of the first two Minkowski functionals may serve as a measure of how crooked the isodensity curves (the surroundings of the excursions sets) are. -- The wandering of the centroids with the density threshold is a natural extension of the well-known centroid shift~\cite{crone}. However, to quantify the amount of substructure present in the cluster it is better to compute the surface content $\zeta$ and the length of the circumference $\eta$ of the triangle formed by the (curvature) centroids. The behaviour of the so defined morphological parameter $\zeta$ can be seen in the fourth panel of Figure \ref{fig:minkclus}.\\
To further condense the detailed information present in the Minkowski functionals we form parameters $\alpha$, $\beta$ and $\gamma$, which overcome the dependence on the density~\cite{vec}.
For example we define $\alpha$ by integrating the deviation of the Euler cha\-racteristic from $1$, which is the value for a highly symmetric reference model. All these parameters are constructed in such a way that they assume the value $0$ if the cluster is spherically symmetric. 
\begin{figure}
\begin{minipage}[t]{.49\linewidth}
\centering
\hspace*{5mm}{\small cg02}\\

\vspace*{0.1cm}
\psfig{file=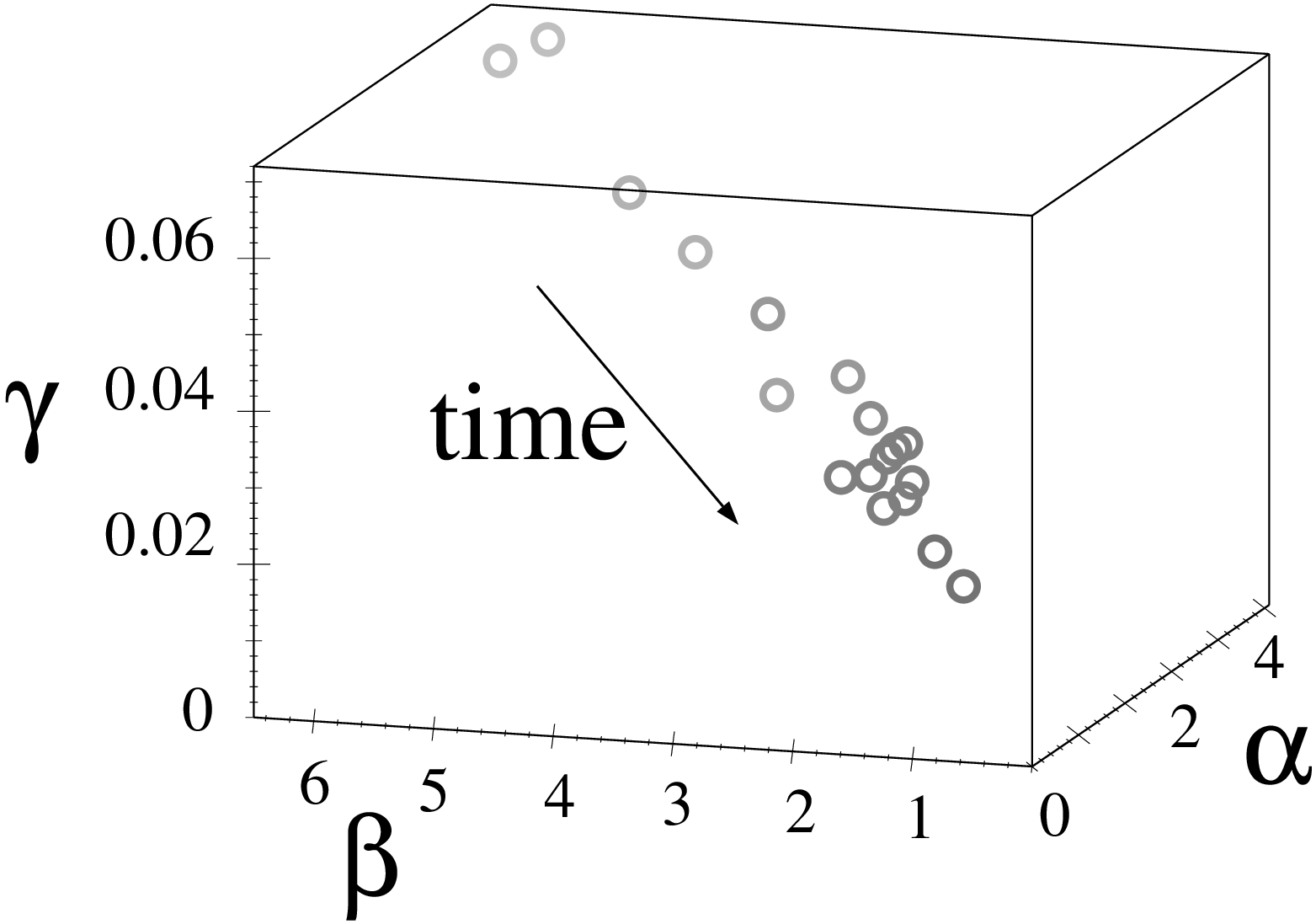,height=2.6cm,width=3.6cm,bbllx=0,bblly=160,bburx=550,bbury=560}
\end{minipage}
\begin{minipage}[t]{.49\linewidth}
\centering
\hspace*{5mm}{\small cg04}\\

\vspace*{0.1cm}
\psfig{file=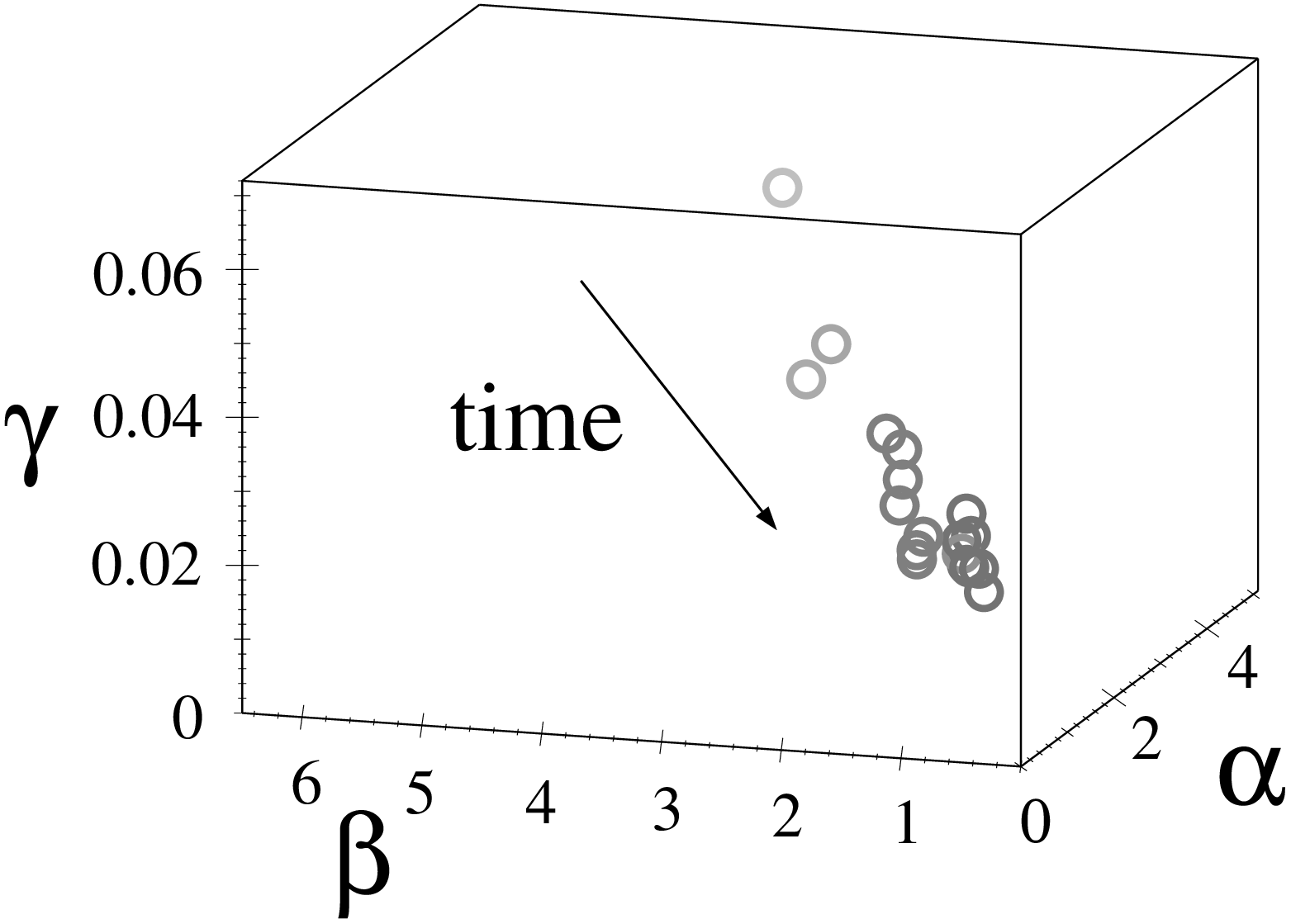,height=2.6cm,width=3.6cm,bbllx=0,bblly=160,bburx=530,bbury=560}
\end{minipage}
\caption{Evolution of simulated clusters ($\Omega_0 = 1$) in the phase diagram}
\label{fig:ev}
\end{figure}
\section{Results and further prospects}
Using the above defined morphological descriptors we can follow the dynami\-cal evolution of simulated~\cite{stein} clusters within a phase-diagram defined in the $\alpha\beta\gamma$-space. Two examples are shown in Figure \ref{fig:ev}. The paths of evolution take  similiar tracks towards a more relaxed, substructure-poor state. Note, however, that these tracks span different scales: cg02 has a higher amount of substructure than cg04. In different cosmogonies the paths will differ; e.g. in low--$\Omega_0$ universes the morphological evolution will stay almost within the $\beta\gamma$-plane. Hence, the cosmological parameters can be determined quantitatively in comparison with observed cluster data. Projection effects in the substructure may be controlled using the stereological properties of the Minkowski functionals.  
\section*{Acknowledgments}
The authors acknowledge support from the {\em Sonderforschungsbereich SFB 375
f\"ur Astroteilchenphysik der Deutschen Forschungsgemeinschaft}. We are indebted to H. Wagner for suggesting these measures and thank M. Bartelmann, M. Kerscher and J. Schmalzing for useful discussions.

\end{document}